\documentclass{article}
\usepackage{spconf,amsmath,amssymb,graphicx,booktabs,tabularx,array,xcolor,tikz,hyperref,flafter}
\usetikzlibrary{arrows.meta,positioning,fit}

\title{QC-MI-OVSEL: BENCHMARKING QUERY-CONDITIONED MULTI-INSTANCE OPEN-VOCABULARY SOUND EVENT LOCALIZATION}

\name{Yongqiang Chen, Peiwei Chang, Yuelan Cheng, Xiatian Zhu, Wenwu Wang}
\address{Centre for Vision, Speech and Signal Processing, University of Surrey, UK\\
\texttt{yongqiang.chen@surrey.ac.uk}}

\begin{document}
\ninept
\maketitle

\begin{abstract}
We study query-conditioned multi-instance open-vocabulary sound event localization (QC-MI-OVSEL), where a system must recover all active physical sources matching a natural-language category query, including when the queried category is absent.
We introduce the QC-MI-OVSEL benchmark, a 40,000-scene first-order Ambisonics benchmark with 80,000 balanced present/absent query views over 83 classes, including 17 novel classes excluded from training and development.
Each frame is annotated with the activity and 3-D direction of every query-matching source, with up to three simultaneous targets.
We further propose Q-RSet, which uses query-conditioned source slots and responsibility-based evidence aggregation for source-set prediction.
Experiments show improved source recovery, localization recall, cardinality estimation, exact-set recovery, and absent-query suppression over complementary baselines.
Analysis indicates that novel categories, paraphrased queries, and higher source multiplicity mainly impair source discovery and set completion, while directional accuracy remains comparatively stable for successfully recovered sources.
\end{abstract}


\section{Introduction}
\label{sec:intro}

Sound event localization and detection (SELD) estimates event activity and direction from multichannel audio \cite{adavanne2019seldnet}.
Multiple Activity-Coupled Cartesian Direction of Arrival (Multi-ACCDOA) extends SELD to multiple simultaneous sources from the same class through
track-wise prediction and permutation-invariant training
\cite{shimada2022multi}, but is restricted to a predefined class set.
Target-conditioned methods instead localize user-specified sounds using predefined class conditions \cite{slizovskaia2022locate} or natural-language
descriptions \cite{zhao2024text}.
Zero-/few-shot and open-vocabulary SELD associate spatial tracks with event representations in a shared language--audio embedding space, enabling
category matching beyond supervised classes
\cite{shimada2024zerofew,shimada2025open}.
More recently, SelectTSL allows multiple target categories to be specified by prompts and selectively localizes the active prompted targets over time
\cite{jiang2026selecttsl}.

However, these formulations do not jointly address the recovery of multiple physical sources that match the same language query.
Such sources may have independent activities and directions, requiring a system to recover the complete query-matching source set, infer its time-varying cardinality, and suppress predictions when the queried category is absent.
Multi-ACCDOA supports same-class multiplicity but assumes a predefined class set \cite{shimada2022multi}, while target-conditioned and open-vocabulary methods primarily focus on localizing queried categories without explicitly modeling complete same-category source sets \cite{zhao2024text,shimada2025open}.
SelectTSL further supports multiple concurrent target sources per prompt with time-varying cardinality \cite{jiang2026selecttsl}, but does not explicitly address scene-level absent queries, generalization to held-out categories, or complete 3-D source-set recovery.

We therefore introduce QC-MI-OVSEL (Query-Conditioned Multi-Instance Open-Vocabulary Sound Event Localization).
Given a four-channel first-order Ambisonics (FOA) scene and a category-level language query, the task is to predict the unordered set of active physical
instances whose category matches the query, together with their 3-D directions.
The accompanying QC-MI-OVSEL benchmark contains 40,000 scenes and 80,000 balanced present/absent query views over 83 classes, including 17 Novel classes excluded from training and development.

We further propose Q-RSet, which uses three learnable source slots to form query-conditioned source hypotheses and gathers source-specific acoustic
evidence through non-exclusive responsibility weighting.
Our contributions are: (1) a query-conditioned multi-instance open-vocabulary sound event localization formulation; (2) the QC-MI-OVSEL benchmark with
explicit absent-query and category-disjoint evaluation; and (3) Q-RSet, a source-set prediction model for recovering multiple query-matching sources.

\section{Task Formulation}
\label{sec:task}

Let $\mathbf{x}\in\mathbb{R}^{4\times N}$ denote an FOA recording and $q$ a
natural-language query associated with category $c_q$.
Each physical source instance $i$ has category $c_i$, framewise activity
$a_{i,t}\in\{0,1\}$, and unit direction
$\mathbf{u}_{i,t}\in\mathbb{S}^{2}$.
The query-matching source set at frame $t$ is
\begin{equation}
\mathcal{G}_t(q)
=
\left\{
(i,\mathbf{u}_{i,t})
\mid
a_{i,t}=1,\;
c_i=c_q
\right\},
\label{eq:targetset}
\end{equation}
with cardinality
$K_t=|\mathcal{G}_t(q)|\in\{0,1,2,3\}$.

\begin{figure}[t]
    \centering
    \includegraphics[width=1.0\columnwidth]{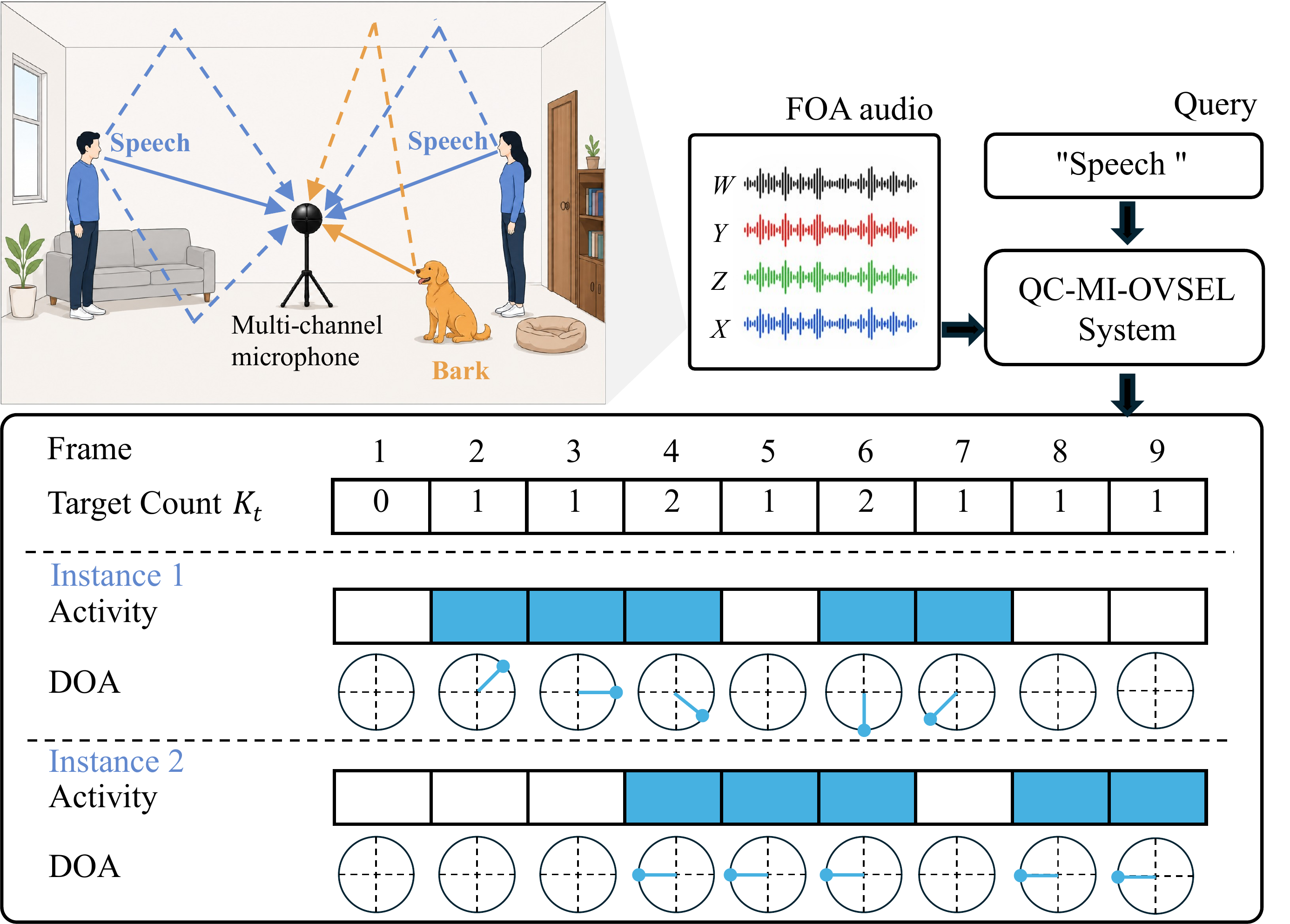}
    \caption{Query-conditioned multi-instance open-vocabulary sound event localization formulation.}
    \label{fig:task_formulation}
\end{figure}

The task is to predict the unordered framewise set
\begin{equation}
(\mathbf{x},q)
\longmapsto
\widehat{\mathcal{G}}_t(q),
\end{equation}
including the empty set when no queried source is active.
As illustrated in Fig.~\ref{fig:task_formulation}, a query may correspond to multiple simultaneously active physical instances, whose activities, directions, and resulting target count vary over time.

A source is treated as a target only if its class matches the queried class.
Canonical queries and paraphrases of the same class share identical reference
targets.

\section{Method}
\label{sec:method}

\begin{figure}[t]
    \centering
    \includegraphics[width=1.0\columnwidth]{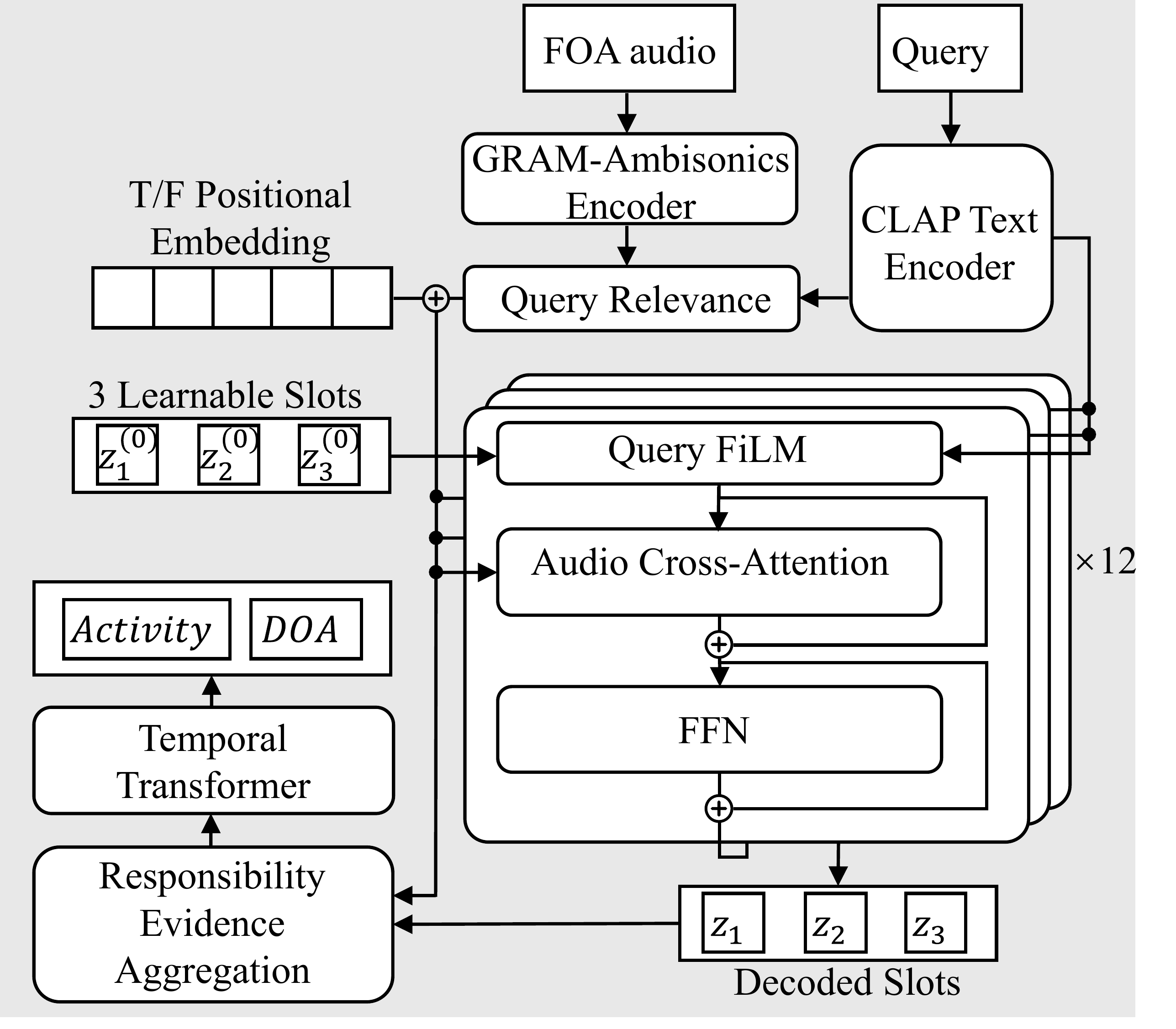}
    \caption{Q-RSet architecture.}
    \label{fig:system}
\end{figure}


We propose Q-RSet, a query-conditioned responsibility-based source-set prediction model that predicts up to three query-matching physical sources with framewise activity and 3-D direction.
It consists of three main stages: query-conditioned acoustic representation, source-hypothesis decoder, and responsibility evidence aggregation. 
These stages progressively identify query-relevant acoustic content, form candidate physical-source hypotheses, and associate each hypothesis with framewise acoustic evidence.
The model structure is presented in Fig.~\ref{fig:system}.

\subsection{Query-Conditioned Acoustic Representation}

A frozen GRAM-Ambisonics encoder \cite{yuksel2026gram} produces a $126\times8$ grid of 768-D acoustic tokens, linearly projected to 256-D.
The query is encoded by a pretrained CLAP text encoder \cite{wu2023large} and projected to the same dimension.
A learned query-relevance module uses each acoustic token $\mathbf{h}_{t,f}$, indexed by time $t$ and frequency $f$, together with the query embedding $\mathbf{q}$ and their element-wise interaction to produce a soft relevance score $r^q_{t,f}\in[0,1]$, where the superscript $q$ indicates conditioning on the query.
The score is learned end-to-end as query-conditioned acoustic relevance rather than an explicitly supervised class probability.
It reweights the acoustic tokens, while temporal and frequency positional embeddings preserve their time--frequency locations.
The resulting representation serves as shared memory for source-hypothesis decoding and subsequent responsibility evidence aggregation.

\subsection{Query-Conditioned Source-Hypothesis Decoder}

Q-RSet uses three learnable source slots to represent up to three simultaneously active query-matching sources.
The slots have independently learned initial embeddings and carry no predefined category, direction, or source identity.
At each decoder layer, the query embedding modulates every slot through feature wise linear modulation (FiLM)~\cite{perez2018film}.
Each conditioned slot then cross-attends to the shared acoustic memory, followed by a feed-forward network with residual connections and layer normalization.
The same query modulation is applied to all slots because they seek different physical instances of the same queried category.
Instance differentiation instead arises from their distinct learned slot states and subsequent cross-attention to the acoustic memory.
Permutation-invariant set supervision then encourages the slots to specialize toward different source hypotheses without imposing a predefined slot identity.

\subsection{Responsibility Evidence Aggregation}
The decoder produces three global source hypotheses, but framewise activity and localization require acoustic evidence specific to each hypothesis. We define the \emph{responsibility} $\rho_{k,t,f}$ as a non-exclusive, slot-specific support score that measures how strongly acoustic token $(t,f)$ provides evidence for source hypothesis $k$.

For slot $k$, its decoded representation $\mathbf{z}_k$ is compared with each query-conditioned acoustic token $\bar{\mathbf{h}}^q_{t,f}$:
\begin{equation}
\rho_{k,t,f}
=
\sigma\!\left(
\frac{
\phi_z(\mathbf{z}_k)^\top
\phi_h(\bar{\mathbf{h}}^q_{t,f})
}{
\sqrt{D}
}
\right),
\label{eq:responsibility}
\end{equation}
where $\sigma(\cdot)$ denotes the sigmoid function, $\phi_z$ and $\phi_h$
are learned projections, $D$ is their output dimension.
Responsibilities are computed independently across slots, allowing the same acoustic token to support multiple source hypotheses.

The query-relevance score $r^q_{t,f}$ indicates whether a token is relevant to the queried class, while $\rho_{k,t,f}$ indicates whether it supports hypothesis $k$.
Their product is used to pool slot-specific frame evidence:
\begin{equation}
\mathbf{g}^{\mathrm{resp}}_{k,t}
=
\frac{
\sum_f r^q_{t,f}\rho_{k,t,f}\bar{\mathbf{h}}^q_{t,f}
}{
\sum_f r^q_{t,f}\rho_{k,t,f}+\epsilon
}.
\label{eq:evidence_pool}
\end{equation}
The pooled evidence is combined with the decoded slot and a global frame-level acoustic feature.
After interpolation to the benchmark frame rate, a shared temporal Transformer models source dynamics, followed by activity and normalized Cartesian 3-D direction of arrival (DOA) heads.

\subsection{Set Matching and Training}

Since source slots are unordered, training uses full-clip permutation-invariant matching based on activity and active-frame DOA errors.
After matching, the model is optimized with binary cross-entropy activity loss and cosine DOA loss on active frames; absent queries contribute only the activity loss.
At inference, slots whose sigmoid activity scores exceed the development-selected threshold are treated as active sources, and their number gives the framewise cardinality.

\section{QC-MI-OVSEL Benchmark}

\textbf{Source preparation.}
We extract non-overlapping event segments of at least 3~s from AudioSet Strong \cite{hershey2021benefit} using the provided onset--offset annotations.
Each segment is resampled to 24~kHz, and used as a single-event source waveform.

\textbf{Ontology and splits.}
The benchmark contains 83 query classes forming an antichain in the ontology, such that no class is an ancestor or descendant of another: 66 base and 17 novel classes.
This avoids ambiguity in exact-category matching caused by selecting both an ontology node and its ancestor or descendant.
Novel classes are excluded from training and development scenes and queries, and appear only at test time.
The 40,000 scenes are split into 30k/2k/8k train/dev/test scenes.
The 8k test scenes are equally divided into four conditions controlling semantic and acoustic shifts:
\emph{Base-ID} uses base queries and training-distribution room impulse response (RIR) rooms;
\emph{Novel-ID} uses novel queries in training-distribution rooms;
\emph{Acoustic-OOD} uses base queries in held-out RIR rooms; and
\emph{Generalized} combines base and novel queries with both seen and held-out rooms.
Source recordings are disjoint across splits.

\textbf{Query views and scene construction.}
Each scene contains one to three instances of its present-query class together with non-target interferers.
Time-varying activity yields frame-level target counts $K_t\in\{0,1,2,3\}$, with static, moving, crossing, close-source, and temporally intermittent
sources.
Each scene provides one present and one absent canonical query, yielding 80,000 balanced query views (60k/4k/16k for train/dev/test).
Each class has eight validated paraphrases used only for linguistic-robustness evaluation.

\textbf{Spatial rendering and annotation.}
Scenes are rendered as 10-s, 24-kHz FOA signals in ACN/SN3D format with channel order $(W,Y,Z,X)$ using measured RIRs from the TAU Spatial RIR Database (TAU-SRIR DB) \cite{politis2022tausrir}.
All sources within a scene use RIRs from the same acoustic environment.
Spatial configurations and reference DOAs are derived from the RIR positions and trajectories used for rendering.
Annotations are provided at 10~Hz and include instance identity, activity, azimuth, and elevation for each query-matching source; DOAs are defined only on active frames.

\section{Experimental Setting}
\label{sec:experimental_setting}

\subsection{Evaluation Metrics}

At each frame $t$, let $\hat K_t$ and $K_t$ denote the numbers of active predicted and reference sources, respectively.
We compute the spherical angular distance between every predicted--reference pair and use the Hungarian algorithm to obtain the optimal one-to-one matching.
Only pairs within $20^\circ$ are considered valid matches.
Let $M_t$ denote the number of valid matches at frame $t$.
Aggregating over frames, we have the true positives (\textbf{TP}), false positives (\textbf{FP}), and false negatives (\textbf{FN}) defined as follows,  
\begin{equation}
\mathrm{TP}=\sum_t M_t,\quad
\mathrm{FP}=\sum_t(\hat K_t-M_t),\quad
\mathrm{FN}=\sum_t(K_t-M_t).
\end{equation}
We measure the query-conditioned source recovery precision in \textbf{F1} score and localization recall (\textbf{LR}) as follows,
\begin{equation}
\mathrm{F1}
=
\frac{2\mathrm{TP}}
{2\mathrm{TP}+\mathrm{FP}+\mathrm{FN}},
\qquad
\mathrm{LR}
=
\frac{\mathrm{TP}}
{\mathrm{TP}+\mathrm{FN}},
\end{equation}
where LE is the mean spherical angular error over the valid matched pairs.
The source-count error is measured by the K-mean absolute error (\textbf{K-MAE}) as
\begin{equation}
\mathrm{K\text{-}MAE}
=
\frac{1}{T}\sum_t |\hat K_t-K_t|.
\end{equation}
\textbf{Exact} is the fraction of active-reference frames for which the complete source set is recovered, requiring $\hat K_t=K_t$ and all matched directions within $20^\circ$. \textbf{Abs-FPR} is the fraction of absent-query scenes that contain any predicted target activity.
Unless otherwise stated, F1, LR, LE, and Exact use a $20^\circ$ localization tolerance.
F1, LR, Exact, and Abs-FPR are reported in percentage, LE in degrees, and K-MAE in the number of sources.

\subsection{Training and Model Selection}

All models are trained and selected using canonical queries only; the paraphrases are reserved for linguistic-robustness evaluation.
We use three random seeds, $\{2027,2028,2029\}$, and train for 40k optimization steps with a batch size of 64.
The trainable parameters are optimized with Adam ($\beta_1=0.9$, $\beta_2=0.999$) and weight decay $10^{-6}$.
The learning rate is warmed up from $2\times10^{-5}$ to $10^{-4}$ over the first 5k steps, cosine-decayed to $5\times10^{-5}$ by 30k, and then kept constant.
The pretrained GRAM \cite{yuksel2026gram} and CLAP \cite{wu2023large} encoders remain frozen. Checkpoints are saved every 5k steps and evaluated on the canonical development set using activity thresholds from 0.05 to 0.95 in increments of 0.05.
For each seed, the checkpoint--threshold pair with the highest development-set F1 is fixed for evaluation. Unless otherwise stated, results are averaged over the three seeds.

\subsection{Baselines}

We compare Q-RSet with four complementary baselines.
\textbf{OVSELD-Adapt} adapts the OpenVocabularySELD model \cite{shimada2025open}.
It performs query-independent SELD followed by semantic filtering of the predicted tracks, which contain activity, 3-D direction, and acoustic event
embeddings matched to the query using frozen CLAP representations.
\textbf{Q-MACCDOA} is a query-conditioned adaptation of Multi-ACCDOA \cite{shimada2022multi}.
Using the same frozen GRAM and CLAP encoders and query-conditioned acoustic representation as Q-RSet, it directly predicts three activity--DOA tracks
without source slots or responsibility aggregation and uses full-clip permutation-invariant matching.
\textbf{TQ-SEL-Adapt} adapts Text-Queried Target Sound Event Localization \cite{zhao2024text} to three activity--DOA tracks with full-clip permutation-invariant training.

\section{Results}
\label{sec:results}

\begin{table}[t]
\caption{Canonical test results on QC-MI-OVSEL benchmark, averaged over three seeds.}
\label{tab:main_results}
\centering
\fontsize{9}{10.5}\selectfont
\setlength{\tabcolsep}{2.1pt}
\begin{tabular}{@{}lcccccc@{}}
\toprule
Model
& F1 $\uparrow$
& LR $\uparrow$
& LE $\downarrow$
& K-MAE $\downarrow$
& Exact $\uparrow$
& Abs-FPR $\downarrow$ \\
\midrule
OVSELD-Adapt
& 47.12 & 37.47 & \textbf{7.93} & 0.47 & 21.02 & 30.01 \\
Q-MACCDOA
& 43.09 & 38.79 & 10.13 & 0.44 & 20.27 & 22.44 \\
TQ-SEL-Adapt
& 26.91 & 37.45 & 11.21 & 0.94 & 13.32 & 96.13 \\
\midrule
\textbf{Q-RSet}
& \textbf{50.36}
& \textbf{46.11}
& 8.86
& \textbf{0.41}
& \textbf{25.89}
& \textbf{20.55} \\
\bottomrule
\end{tabular}
\end{table}

\begin{table}[t]
\caption{Generalization and language robustness of Q-RSet.}
\label{tab:generalization}
\centering
\fontsize{9}{10.5}\selectfont
\setlength{\tabcolsep}{2.0pt}
\begin{tabular}{@{}lcccccc@{}}
\toprule
Condition
& F1 $\uparrow$
& LR $\uparrow$
& LE $\downarrow$
& K-MAE $\downarrow$
& Exact $\uparrow$
& Abs-FPR $\downarrow$ \\
\midrule
\multicolumn{7}{l}{\emph{Generalization}} \\
Base-ID      & 57.07 & 56.56 & 8.74 & 0.33 & 31.84 & 18.15 \\
Novel-ID     & 34.27 & 26.90 & 8.84 & 0.60 & 16.65 & 28.32 \\
Acoustic-OOD & 56.92 & 55.94 & 8.79 & 0.32 & 31.82 & 17.48 \\
Generalized  & 49.87 & 45.12 & 9.10 & 0.40 & 23.29 & 18.23 \\
\midrule
\multicolumn{7}{l}{\emph{Language robustness}} \\
Canonical    & 50.36 & 46.11 & 8.86 & 0.41 & 25.89 & 20.55 \\
Paraphrase   & 37.58 & 30.36 & 9.10 & 0.55 & 17.43 & 23.11 \\
\bottomrule
\end{tabular}
\end{table}

\begin{table}[t]
\caption{Q-RSet performance by maximum simultaneously active target count.}
\label{tab:kmax}
\centering
\fontsize{9}{10.5}\selectfont
\setlength{\tabcolsep}{3.0pt}
\begin{tabular}{@{}cccccc@{}}
\toprule
$K_{\max}$ & F1 $\uparrow$ & LR $\uparrow$ & LE $\downarrow$
& K-MAE $\downarrow$ & Exact $\uparrow$ \\
\midrule
1 & 56.04 & 52.14 & 8.16 & 0.43 & 39.27 \\
2 & 52.72 & 44.23 & 8.90 & 0.69 & 23.25 \\
3 & 54.80 & 46.17 & 9.22 & 0.89 & 14.89 \\
\bottomrule
\end{tabular}
\end{table}

\begin{table}[t]
\caption{
Architecture analysis and component ablations of Q-RSet.
}
\label{tab:ablation}
\centering
\fontsize{9}{10}\selectfont
\setlength{\tabcolsep}{1.8pt}
\renewcommand{\arraystretch}{0.92}
\begin{tabular}{@{}lcccccc@{}}
\toprule
Model
& F1 $\uparrow$
& LR $\uparrow$
& LE $\downarrow$
& K-MAE $\downarrow$
& Exact $\uparrow$
& Abs-FPR $\downarrow$ \\
\midrule

\multicolumn{7}{l}{\emph{Decoder depth}} \\
3 layers
& 47.23 & 43.45 & 9.26 & 0.44 & 22.41 & 20.58 \\
6 layers
& 49.28 & 44.87 & 9.00 & 0.42 & 24.75 & 20.88 \\
12 layers
& \textbf{50.36} & \textbf{46.11} & \textbf{8.86}
& \textbf{0.41} & \textbf{25.89} & \textbf{20.55} \\

\midrule
\multicolumn{7}{l}{\emph{Component ablations}} \\
Q-RSet
& \textbf{50.36} & \textbf{46.11} & \textbf{8.86}
& \textbf{0.41} & \textbf{25.89} & 20.55 \\
w/o Relevance
& 49.19 & 45.63 & 9.12 & 0.42 & 24.41 & 20.25 \\
w/o Query FiLM
& 48.04 & 43.88 & 9.22 & 0.43 & 23.58 & \textbf{19.49} \\
w/o Cross-Attn
& 46.34 & 42.41 & 9.65 & 0.42 & 21.86 & 22.18 \\
w/o Responsibility
& 48.22 & 44.38 & 9.34 & \textbf{0.41} & 24.58 & 21.56 \\
w/o Temporal
& 44.08 & 40.40 & 9.63 & 0.43 & 20.02 & 22.00 \\
\bottomrule
\end{tabular}
\end{table}


\subsection{Main Comparison}

Table~\ref{tab:main_results} shows that Q-RSet achieves the strongest overall source-set recovery, obtaining the best F1, LR, K-MAE, Exact, and Abs-FPR.
Compared with OVSELD-Adapt, Q-RSet improves F1 by 3.24 percentage points and LR by 8.64 points, while reducing K-MAE from 0.47 to 0.41 and Abs-FPR from 30.01\% to 20.55\%. Exact also improves from 21.02\% to 25.89\%, indicating better complete-set recovery.
OVSELD-Adapt nevertheless achieves a lower LE
($7.93^\circ$ versus $8.86^\circ$), indicating higher directional precision among successfully recovered sources.
The main advantage of Q-RSet therefore lies in recovering a more complete query-matching source set rather than improving the angular precision of already recovered sources.

The controlled comparison with Q-MACCDOA further supports the proposed set-prediction formulation. Using the same query-conditioned acoustic representation, direct multi-track regression achieves 43.09\% F1, whereas Q-RSet reaches 50.36\% and improves LR, LE, K-MAE, Exact, and Abs-FPR.
This supports source-hypothesis decoding and source-specific evidence attribution over direct query-conditioned multi-track regression.
TQ-SEL-Adapt performs particularly poorly on absent queries, with an Abs-FPR of 96.13\%, indicating that extending target-queried localization to multiple tracks does not provide reliable query-absence handling.

\subsection{Generalization and Language Robustness}

The generalization results show that semantic shift is more challenging than the acoustic shift considered here.
Compared with Base-ID, Novel-ID reduces F1 from 57.07\% to 34.27\% and LR from 56.56\% to 26.90\%, while K-MAE increases from 0.33 to 0.60.
In contrast, Acoustic-OOD remains close to Base-ID.
LE varies only from $8.74^\circ$ to $9.10^\circ$ across the four conditions, suggesting that the main generalization bottleneck is source discovery and count recovery rather than directional estimation.

Paraphrasing shows a similar pattern.
F1 decreases from 50.36\% to 37.58\%, retaining 74.6\% of canonical performance, and LR from 46.11\% to 30.36\%, while K-MAE increases from 0.41 to 0.55.
However, LE changes by only $0.24^\circ$.
Linguistic variation mainly affects identifying and counting queried sources, with comparatively little effect on the localization accuracy of successfully recovered sources.

\subsection{Multi-Instance Analysis}

Table~\ref{tab:kmax} evaluates present queries according to the maximum number
of simultaneously active query-matching sources.
The clearest degradation with increasing multiplicity occurs in cardinality
and complete-set recovery: K-MAE increases from 0.43 at $K_{\max}=1$ to
0.89 at $K_{\max}=3$, while Exact decreases from 39.27\% to 14.89\%.
In comparison, F1 remains between 52.72\% and 56.04\%, and LE increases moderately from $8.16^\circ$ to $9.22^\circ$. Notably, F1 does not decrease monotonically with $K_{\max}$, whereas K-MAE and Exact show clear monotonic degradation.
This suggests that increasing multiplicity primarily challenges set completeness rather than individual-source recoverability.
Q-RSet can therefore recover individual sources even in multi-instance
scenes, but recovering the correct number and the complete simultaneous
source set becomes progressively more difficult.
The three subsets contain 2,458, 4,029, and 1,513 present-query views,
respectively.

\subsection{Ablation Study}

Increasing the source-hypothesis decoder from 3 to 6 and 12 layers improves F1 from 47.23\% to 49.28\% and 50.36\%, respectively.
The 12-layer model also performs best in LR, LE, K-MAE, Exact, and Abs-FPR, motivating its use in the final Q-RSet.

Removing every component reduces F1 consistently across all three matched seeds.
The largest mean degradation occurs without the Temporal Transformer ($-6.27$ percentage points), followed by Audio Cross-Attention ($-4.01$ points).
Removing Query FiLM, Responsibility, and Query Relevance decreases F1 by 2.32, 2.14, and 1.17 points, respectively.
These results support query conditioning at both the acoustic-representation
and source-hypothesis stages, scene-adaptive acoustic interaction,
hypothesis-specific evidence attribution, and temporal modeling.
Some removals slightly improve isolated secondary metrics, such as Abs-FPR
or K-MAE, but none improves overall source recovery.

\section{Conclusion}
We introduced QC-MI-OVSEL, together with the QC-MI-OVSEL benchmark and Q-RSet for query-conditioned multi-instance open-vocabulary sound event localization.
Q-RSet improves source recovery, cardinality estimation, complete-set recovery, and absent-query suppression over complementary baselines.
Across novel categories, paraphrased queries, and increasing source multiplicity, performance degradation arises mainly from source discovery, counting, and set completion, while directional accuracy remains comparatively stable for successfully recovered sources.
These results highlight query-conditioned source-set recovery, rather than DOA estimation alone, as a central challenge for open-vocabulary multi-instance spatial audio understanding.

\bibliographystyle{IEEEbib}
\bibliography{references}

\end{document}